\documentclass[]{emulateapj}
\usepackage{epsfig, natbib, graphicx, color}
\usepackage{apjfonts}

\input epsf

\newcommand{\Mpc}{\mbox{Mpc}}
\newcommand{\hMpc}{h^{-1}\,\mathrm{Mpc}}
\newcommand{\msun}{M_\odot}
\newcommand{\bm}[1]{\mathbf{#1}}

\newcommand{\Nobs}{N_{200}}

\newcommand{\avg}[1]{\left\langle #1 \right\rangle}

\newcommand{\lk}{{\cal{L}}}

\newcommand{\bx}{\bm{x}}

\newcommand{\Rc}{R_c}

\newcommand{\kpc}{h^{-1}\,\mathrm{kpc}}

\newcommand{\Lx}{L_X}
\newcommand{\lnl}{\ln \Lx}

\shortauthors{ROZO ET AL.}
\shorttitle{Improved Richness Estimates}

\begin{document}
\title{An Improved Cluster Richness Estimator}
\author{Eduardo Rozo\altaffilmark{1}, Eli S. Rykoff\altaffilmark{2}, Benjamin P. Koester\altaffilmark{3,4}, Timothy McKay\altaffilmark{5,6,7}, 
Jiangang Hao\altaffilmark{5}, August Evrard\altaffilmark{5,6,7}, Risa H. Wechsler\altaffilmark{8}, Sarah Hansen\altaffilmark{3,4}, 
Erin Sheldon\altaffilmark{9}, David Johnston\altaffilmark{10}, Matthew Becker\altaffilmark{3,4}, James Annis\altaffilmark{11}, 
Lindsey Bleem\altaffilmark{3}, Ryan Scranton\altaffilmark{12}}

\altaffiltext{1}{Center for Cosmology and Astro-Particle Physics (CCAPP), The Ohio State University, Columbus, OH 43210}
\altaffiltext{2}{TABASGO Fellow, Physics Department, University of California
  at Santa Barbara, 2233B Broida Hall, Santa Barbara, CA 93106}
\altaffiltext{3}{Department of Astronomy and Astrophysics, The University of Chicago, Chicago, IL 60637}
\altaffiltext{4}{Kavli Institute for Cosmological Physics, The University of Chicago, Chicago, IL 60637} 
\altaffiltext{5}{Physics Department, University of Michigan, Ann Arbor, MI 48109}
\altaffiltext{6}{Astronomy Department, University of Michigan, Ann Arbor, MI 48109}
\altaffiltext{7}{Michigan Center for Theoretical Physics, Ann Arbor, MI 48109}
\altaffiltext{8}{Kavli Institute for Particle Astrophysics \& Cosmology,
  Physics Department, and Stanford Linear Accelerator Center,
  Stanford University, Stanford, CA 94305}
\altaffiltext{9}{Center for Cosmology and Particle Physics, Physics Department, New York University, New York, NY 10003}
\altaffiltext{10}{Jet Propulsion Laboratory, 4800 Oak Grove Drive, Pasadena, CA 91109}
\altaffiltext{11}{Fermi National Accelerator Laboratory, P.O. Box 500, Batavia, IL 60510}
\altaffiltext{12}{Department of Physics and Astronomy, Universityof Pittsburgh, 3941 O'Hara St., Pittsburgh, PA 15260}

\begin{abstract}
Minimizing the scatter between cluster mass and accessible observables
is an important goal for cluster cosmology. In this work, we introduce a new matched filter richness estimator,
and test its performance using the maxBCG cluster catalog.  Our new estimator
significantly reduces the variance in the $\Lx-$richness relation, from $\sigma_{\lnl}^2=(0.86\pm0.02)^2$ to 
$\sigma_{\lnl}^2=(0.69\pm0.02)^2$.  Relative to the maxBCG richness estimate, it also removes the strong redshift 
dependence of the richness scaling relations, and is significantly more robust to
photometric and redshift errors.  These improvements are largely due to our more sophisticated treatment
of galaxy color data.  We also demonstrate the scatter in the $\Lx-$richness relation depends on the
aperture used to estimate cluster richness, and introduce a novel approach for optimizing said aperture which
can be easily generalized to other mass tracers.
\end{abstract}


\section{Introduction}

The dependence of the halo mass function on cosmology is a problem that is well understood
both analytically \citep{pressschechter74,bondetal91,shethtormen02} and numerically 
\citep{jenkinsetal01,warrenetal06,tinkeretal08}.  In principle, this detailed understanding allows one
to place tight constraints on the amplitude of the primordial power spectrum and on
dark energy parameters \citep[e.g.][]{holderetal01,haimanetal01}.  In practice, life is not so simple.
Cluster mass is not an observable, and so we must rely on other quantities that
trace mass to estimate the halo mass function.  In this context, 
observables that are tightly correlated with mass and whose scatter is well understood are highly
desirable, as they permit a more accurate measurement of the mass function.

One such mass tracer, and the subject of interest for this work, is the so called cluster richness, 
a measure of the galaxy content of a cluster.  Relative to other popular mass 
tracers such as X-ray properties, SZ-decrements, and galaxy velocity dispersion, optical richness 
has unique advantages and disadvantages. Its unique advantages are:
\begin{enumerate}
\item cluster richness can be easily estimated with inexpensive, photometric optical data.
\item cluster richness can be estimated for both massive clusters and low mass groups.
\end{enumerate}
The first of these two properties is significant because it implies that cluster richness estimates are readily available
given any large, photometric optical survey such as the SDSS \citep{yorketal00} , DES\footnote{http://www.darkenergysurvey.org/},
or LSST\footnote{http://www.lsst.org/lsst\_home.shtml}.
The latter property, on the other hand, is an important advantage for a much more interesting reason.  

Beginning with \citet{whiteetal93},
cosmological constraints from galaxy clusters have been presented as a degeneracy relation
$\sigma_8\Omega_m^{\gamma}=constant$ where $\gamma\approx 0.5$, $\sigma_8$ is a parameter specifying the amplitude
of the primordial power spectrum, and $\Omega_m$ is the matter density of the universe in units of the critical density. 
The existence of this degeneracy is easy to explain~\citep[][]{rozoetal04}: suppose that we only measured the abundance
of galaxy clusters at a single mass scale.  Since the halo mass function depends on both $\sigma_8$ and $\Omega_m$,
it is evident that with just one observable there must be a degeneracy between these two parameters.  But what if
we measure the halo mass function over a range of scales?  This is roughly equivalent to measuring the
amplitude and slope of the halo mass function at the statistical pivot point.  If the mass range probed is small, then the
slope of the mass function is not well constrained, and the degeneracy between $\sigma_8$ and $\Omega_m$ will
remain.  In order to break this degeneracy, a measurement of the halo mass function over a large range of masses 
is necessary. Currently, only spectroscopic velocity measurements and optical richness estimates can probe a mass 
range wide enough to successfully break this degeneracy, but the former requires considerably more observing resources.

There are, however, important disadvantages to using cluster richness as a mass tracer. For instance, historically, the 
fact that the relation between cluster richness and mass cannot be predicted a priori based on simple physical arguments
was viewed as a significant drawback.  Nowadays, however, this argument holds little sway, since the level of accuracy required
for precision comsology in our a priori knowledge of cluster scaling relations
is pushing current research towards a self-calibrating approach, in which both cosmology
and cluster scaling relations are simultaneously constrained from the data \citep[][]{limahu04,majumdaretal04,
limahu05,hucohn06,wuetal08}.  Thus, in so far as self-calibration is necessary to insure one-self against possible
biases in cosmological estimates, the lack of a simple physical model for predicting cluster richness is no longer a serious
drawback.

Another reason why optical richness estimates fell out of favor relative to other mass tracers is that, in the past,
richness estimates were known to suffer from significant projection effects, which resulted in impure cluster
samples as well as large scatter in the mass-richness relation.  
Abell made one of the first systematic attempts at measuring richness \citep{abell58,abell89} in defining his richness classes. 
He tried to minimize projection by only counting galaxies dimmer than $m_3$, the magnitude of the third brightest
cluster galaxy, but brighter than $m_3+2$. The bright cut is aimed at foreground interlopers, while the dim cut
reduces the contribution of the galaxy background. Later methods 
used similar counting techniques but included a proper account of the background \citep[e.g.][]{bahcall81} . 
Since then, more sophisticated algorithms have been developed and applied to CCD-based 
imaging \citep[e.g. matched-filter methods][]{postmanetal96,bramel00,yee99,kochaneketal03,dongetal07}.

Projection effects are now a much more benign problem thanks to these more sophisticated richness measurement 
techniques, the advent of accurate photometric data enabled by modern CCDs, and most recently, the well-known observations 
that ellipticals and cluster E/S0 galaxies in particular tend to form a tight ridgeline 
in color-magnitude space \citep{visvanathan77,boweretal92,gladdersyee00,kmawe07a}.  
This color clustering has been integral to richness measurements in the SDSS \citep{gotoetal02,milleretal05,kmawe07a} 
and the Red Sequence Cluster
Survey  \citep[RCS:][]{gladders05}, and such color-based measures have been shown to be effective mass
tracers \citep{yee03,muzzin07,sheldonetal07,johnstonetal07,rmbej08,beckeretal07}.

While richness estimates show a strong correlation with other mass proxies \citep[e.g.][]{yee03,daietal07,sheldonetal07,johnstonetal07,beckeretal07,rembj08}, 
considerable scatter 
in the mass--richness relation still remains. For instance, the richness measure used in the RCS cluster catalog
has a logarithmic scatter of $\sigma_{\ln M}\approx 0.8$ \citep{gladdersetal07}, while for maxBCG clusters
the number is closer to $\sigma_{\ln M}\approx 0.5$~\citep{rozoetal08a}.
This is to be compared to the scatter for X-ray mass tracers,
which is expected to be as low as $\approx 8\%$ for $Y_X$ based on simulations \citep{kravtsovetal06}, 
or as high as $\approx 25\%$ for non-core
extracted soft X-ray band luminosities~\citep[e.g.][]{sebsn06,vbefh08}. Clearly, much improvement is needed to 
bring the scatter of richness measures to the level of X-ray mass tracers.

This work is aimed at reducing the variance in the richness-mass relation.  We do this by explicitly constructing a new 
richness estimator that significantly reduces the scatter in mass at fixed richness for maxBCG clusters. Relative to $\Nobs$ of maxBCG,
we introduce two significant differences.   The first of these involves using a matched filter algorithm
to estimate cluster richness.  Matched filters have been used in the literature before \citep{postmanetal96,kochaneketal03}.
Unlike those works, however, our matched filter includes a color component, which is of critical importance for reducing
projection effects over the redshift range spanned by our cluster sample.\footnote{In
\citet{kochaneketal03}, the low redshift of the clusters make single band magnitudes
better proxies for distance than colors, so the lack of a color filter in the richness estimator is less important
for their work.}
In that sense, our filter is closer in spirit to that of \citet{dongetal07}, who include a photometric redshift filter into 
their richness estimate.  We also note here that group-scale 
studies suggest 
that some measure of the average color in the cluster is indicative of mass, particularly below 
$ \sim 10^{14} M_{\odot}$ \citep[][]{martinez02,martinez06,weinmann06,hansenetal07} .  

The second difference we introduce is the way in which the aperture used to estimate cluster richness is determined.
Generically, cluster richness estimators involve counting the number of galaxies within some specified aperture, which
can thus be interpreted as defining the ``size'' of the cluster.  This begs the question, then, of how is one to select the
correct size of a cluster a priori?  Theoretically, halo sizes are usually defined
in terms of $R_\Delta$, a radius which encompasses a mean density that is $\Delta$ times either the mean or the
critical density of the universe (conventions vary from author to author).  Unfortunately, not only is such a definition
not applicable observationally, authors vary both on the reference background density (critical versus mean mass density),
and on the specific overdensity value.  Thus, even though significant progress has been made \citep{cuestaetal08}, a 
definitive definition of halo size remains elusive.  

In this work, we approach this question with observations in mind.  That is, rather than coming with a preconceived
notion of what the radius of a cluster is, we let the data tell us what the optimal radii for our clusters is by demanding that optical richness
be as tightly correlated as possible with X-ray luminosity.
The idea is as follows: first, one posits a scaling relation between cluster
richness and cluster radius.  When estimating cluster richness, one then demands that the richness-radius scaling relation be
satisfied.  For instance, given a cluster, one can simply make an initial guess for its richness.  Using the richness-radius scaling
relation, one can then draw a circle of the appropriate radius, and count the number of galaxies within it.  If the richness was underestimated,
one will find too many galaxies, signaling that the richness estimate must be increased.  Proceeding in this way, one can quickly 
zero in on the appropriate richness for the object.

This does, however, leave open the question of what the correct richness-radius relation is.  
Since we are interested in finding a new richness estimator that is tightly
correlated with halo mass, we can use the scatter in the mass--richness relation as our figure of merit to determine the
``correct'' richness-radius relation.  In practice, we use the
$\Lx-$richness scatter rather than the mass--richness scatter because the scatter in mass is not directly observable.
We emphasize that since the mass scatter at fixed X-ray luminosity \citep[see e.g.][]{vikhlininetal08} is considerably tighter than
the corresponding scatter at fixed richness~\citep{rozoetal08a}, the use of X-ray luminosity as a mass tracer for
our purposes is well justified.

The layout of the paper is as follows. We describe the data sets used in this work in \S~\ref{sec:data}.  Our
matched filter estimator is introduced in \S~\ref{sec:matchfilter}, followed by
our method for determining the optimal radius-richness relation in
\S~\ref{sec:methods}.  We present our results in section \S~\ref{sec:results}.  In investigating the properties
of our new richness measure, we have
discovered that the redshift evolution of the richness-mass relation of our new
estimator is much more mild than that measured for $N_{200}$.  
These results and the corresponding discussion are presented in
\S~\ref{sec:zdep}.  We summarize our results
and present our conclusions in \S~\ref{sec:conclusions}.  Throughout, whenever needed a flat $\Lambda$CDM
cosmology with $\Omega_M=0.3$ and $h=1.0$ was assumed.


\section{Data}
\label{sec:data}

The data for the analysis presented in this work comes from two large area
surveys, the Sloan Digital Sky Survey~\citep[SDSS: ][]{york00} and the ROSAT
All-Sky Survey~\citep[RASS: ][]{vogesetal99}.  SDSS imaging data are used to
select clusters and to measure their matched filter richness; RASS data provide
0.1-2.4 keV X-ray fluxes, which we convert into estimates of the X-ray luminosity
of the clusters.

\subsection{SDSS}

The imaging and spectroscopic surveys that comprise the SDSS are currently in
the sixth Data Release \citep{adelman08}. This release includes nearly 8500
square degrees of drift-scan imaging in the the Northern Galactic Cap, and
another 7500 square degrees of spectroscopic observations of stars, galaxies,
and quasars.

The camera design \citep{gunn06} and drift-scan imaging strategy of the SDSS
enable acquisition of nearly simultaneous observations in the $u,g,r,i,z$
filter system \citep{fukugita96}. Calibration \citep{hogg01,smith02,tucker06},
astrometric \citep{pier03}, and photometric \citep{lupton01} pipelines reduce
the data into object catalogs containing a host of measured parameters for each
object.

The maxBCG cluster sample and the galaxy catalogs used to remeasure cluster
richness in this paper are derived from the SDSS.  The galaxy
catalogs are drawn from an area approximately coincident with DR4
\citep{adelman06}. Galaxies are selected from SDSS object catalogs as described
in \citep{sheldon07}. In this work we use $\tt{CMODEL\_COUNTS}$ as our total
magnitudes, and $\tt{MODEL\_COUNTS}$ when computing colors. Bright stars, survey edges 
and regions of poor seeing are masked as previously described \citep{kmawe07b,sheldon07}.

\subsection{Cluster Sample}

We obtain sky locations, redshift estimates, and initial richness values from
the maxBCG cluster catalog.  Details of the selection algorithm and catalog
properties are published elsewhere~\citep{kmawe07b,kmawe07a}.  In brief, maxBCG
selection relies on the observation that the galaxy population of rich clusters
is dominated by luminous, red galaxies clustered tightly in color (the E/S0
ridgeline).  Since these galaxies have old, passively evolving stellar
populations, their $g-r$ color closely reflects their redshift.  The brightest
such red galaxy, typically located at the peak of the galaxy density, defines the
cluster center.

The maxBCG catalog is approximately volume limited in the redshift range $0.1
\le z \le 0.3$, with very accurate photometric redshifts ($\delta{}z \sim
0.01$).  Studies of the maxBCG algorithm applied to mock SDSS catalogs indicate
that the completeness and purity are very high, above $90\%$~\citep{kmawe07b,
rwkme07a}.  The maxBCG catalog has been used to investigate the scaling of
galaxy velocity dispersion with cluster richness~\citep{bmkwr07} and to derive
constraints on the power spectrum normalization, $\sigma_8$, from cluster
number counts~\citep{rwkme07a}.

The primary richness estimator used in the maxBCG catalog is $N_{200}$, defined
as the number of galaxies with $g-r$ colors within $2\sigma$ of the E/S0
ridgeline as defined by the BCG color, brighter than $0.4\,L_{*}$ (in
$i$-band), and found within $r_{200}^{gal}$ of the cluster center. 
$r_{200}^{gal}$ is a cluster radius that depends upon the number of galaxies
within a fixed aperture $1\ h^{-1}\ \Mpc$ of the BCG, labeled $N_{gals}$,
with the relation $r_{200}^{gal}(N_{gals})$ being calibrated so that, on average,
the galaxy overdensity within $r_{200}^{gal}$ is
$200\Omega_m^{-1}$ assuming $\Omega_m=0.3$ \citep{hmwas05}.
The full catalog comprises $13,823$ objects with a richness threshold
$\Nobs\geq 10$, corresponding to $M\gtrsim 5\cdot 10^{13}\ h^{-1}\ \msun$ \citep{johnstonetal07}.

As mentioned in the introduction, we re-estimate the cluster richness for every object
in the maxBCG catalog, and measure the corresponding scatter in the $\Lx-$richness relation.
When doing so, we always limit ourself to the 2000 richest clusters, ranked according to
the new richness estimate.   This cut is made to ensure that our results are insensitive
to the $N_{200}\geq 10$ cut of the maxBCG catalog.  That is, the number of clusters
with $N_{200}\geq 10$ that fall within the 2000 richest clusters for any of the new richness
measures considered has no impact on the recovered scatter.  We also note that
our choice of always selecting the 2000 richest clusters also implies that the
specific cluster sample used to estimate the scatter in the $\Lx-$richness relation varies
somewhat as we vary the richness estimator.

\subsection{X-ray Measurements}
\label{sec:xray}

The scatter in $\Lx$ at fixed richness is estimated using a slight variant of
the method presented in \citet{rmbej08}.  Briefly, we use the RASS photon maps
to estimate the 0.5-2.0~keV X-ray flux at the location of each cluster, which
is used to derive $\Lx$~[0.1-2.4 keV] using the cluster photometric redshift
\citep[the conversion factors are similar to those used in][]{bsgcv04}.  We
then perform a Bayesian linear least squares fit to $\ln L_x$ as a function of
$\ln N$, where $N$ is the richness parameter to be tested.  The variance in
$\ln \Lx$ is included as a free parameter.  The fit is done following the
algorithm presented in \citet{k07}, and correctly takes into account upper
limits for $\Lx$ for those clusters with upper limits on X-ray emission.

It is important to note here that the estimated X-ray luminosity of a cluster
depends on the aperture used to measure $\Lx$.  \citet{rmbej08} used a fixed
$750\ \kpc$ aperture as a compromise between needing a large aperture to avoid
losing X-ray photons due to the ROSAT PSF and cluster miscentering, and the
need for a small aperture in order to increase the signal to noise of the
cluster emission.  Further work has shown that the scatter in $\Lx$ at fixed
$N_{200}$ is minimized when using an aperture of $1\ \hMpc$.  The corresponding
scatter for the top 2000 maxBCG clusters is $\sigma_{\lnl|\Nobs}=0.96\pm
0.03$.\footnote{The attentive reader will note that the quoted scatter in $\Lx$
at fixed richness is significantly larger than the scatter in mass at fixed
richness quoted in the introduction, which was closer to $0.5$.  Given a slope
of $\approx 1.6$ in the $\Lx-M$ relation, a scatter of $0.96$ in $\Lx$
corresponds to $\approx 0.96/1.6 \approx 0.6$ scatter in mass.  The remaining
$10\%$ difference is because the scatter in \citet{rozoetal08a} uses
the scatter of the 1000 richest clusters, which is smaller than that of the
2000 richest clusters by 0.1.}

The nature of the present exercise has the benefit of assigning a cluster
radius $R_c$, to each individual cluster, so it is natural to measure $\Lx$ in
the same scale as the optical richness.  Thus, in this work, we estimate $\Lx$
using a variable aperture which depends upon the cluster's richness.  Using 
a fixed $1\,\hMpc$ aperture to estimate $\Lx$  does not have a large
effect on our results, for reasons that will be discussed below.  Finally, we
note that very small physical apertures are impractical for the most distant
clusters due to the large size of the RASS PSF, which corresponds to a physical
scale of $300\,\kpc$ (FWHM) at $z=0.23$, the median redshift of the maxBCG
catalog.  Therefore, we place a fixed minimum aperture of $500\,\kpc$ for each
cluster.  We discuss the small effect of this aperture cutoff in
\S~\ref{sec:methods}.  

\subsection{\label{sec:cleaning}Cleaning the Sample}

Our analysis depends on a combination of optical and X-ray measurements of
maxBCG clusters using SDSS and RASS data.  As discussed in detail in \citet[see
\S~5.6]{rmbej08}, there is clear evidence that cool core clusters
increase the scatter in X-ray cluster properties.  High resolution X-ray
imaging of clusters allows the exclusion of cluster cores, reducing the scatter
in observed X-ray properties~\citep[e.g.][]{ombe06,crbiz07,m07}.
Unfortunately, the broad PSF of RASS means that it is impossible to exclude the
cores of clusters in this work.  In order to asses how robust our results are
to the presence or absence of cooling flow clusters in the cluster sample, we
have created a ``clean'' sample of maxBCG clusters by removing all known cool
core clusters that might have boosted global X-ray luminosity and may
significantly bias our results.  In addition, we have removed apparently X-ray
bright maxBCG clusters that were determined via inspection to have their X-ray
flux significantly contaminated by foreground objects such as stars, low
redshift galaxy clusters, and AGN.

There does not exist a complete, unbiased catalog of cool core X-ray clusters.
The presently described cleaning procedure is not intended to be complete, and
is intended only to give some sense of the robustness of our results to the presence
of cooling flow clusters.  Following \citet{rmbej08}, we have
assembled all the known cool core clusters from the literature.  This includes:
A750, A1835, Z2701, Z3146, Z7160, RXC~2129.6+0005~\citep{bfsaj05},
A1413~\citep{crbiz07}, A2244~\citep{pfeaj98}, and
RXC~J1504.1$-$0248~\citep{bbzsn05}.  From here on, the maxBCG catalog presented
in \citet{kmawe07a} is referred to as the ``full'' cluster sample, and
the subsample described above is referred to as the ``clean'' cluster sample.


\section{Matched Filter Richness Estimators}
\label{sec:matchfilter}

\subsection{Derivation of the Matched Filter Richness Estimator}

Let $\bm{x}$ be a vector characterizing the observable properties of a galaxy (e.g. galaxy color
and magnitude). We model
the projected galaxy distribution around clusters as a sum $S(\bx)=\lambda u(\bx|\lambda)+b(\bx)$
where $\lambda$ is the number of cluster galaxies,
$u(\bx|\lambda)$ is the cluster's galaxy density profile normalized
to unity, and $b(\bx)$ is density of background (i.e. non-member) galaxies.
The probability that a galaxy found near a cluster is actually a cluster member is given by
\begin{equation}
p(\bx) = \frac{\lambda u(\bx|\lambda)}{\lambda u(\bx|\lambda)+b(\bx)}.
\end{equation}
Consequently, the total number of cluster galaxies $\lambda$ must satisfy the constraint equation
\begin{equation}
\lambda  = \sum p(\bx|\lambda) = \sum \frac{\lambda u(\bx|\lambda)}{\lambda u(\bx|\lambda)+b(\bx)}
\label{eq:mfrichness}
\end{equation}
where the sum is over all galaxies in the cluster field.  If the filters $u(\bx|\lambda)$ and $b(\bx)$
are known, then given an observed galaxy distribution
$\{\bx_1, ..., \bx_N\}$ around a cluster we can define a richness estimator $\hat\lambda$ as the
solution to equation \ref{eq:mfrichness}.  As it turns out, one can also derive this expression
using a maximum likelihood approach.  Interested readers are referred to appendix
\ref{app:maxlkhd} for details.  From now on, the letter $\lambda$ shall always refer to a matched
filter richness estimate obtained with equation \ref{eq:mfrichness}.


\subsection{Cluster Radii and Matched Filter Richness Estimates}

Consider again Eqn.~\ref{eq:mfrichness}.  As mentioned before, the sum used in Eqn.~\ref{eq:mfrichness} needs to extend over
all galaxies.  In practice, of course, one needs to add over all galaxies within some cutoff radius $R_c$.  Operationally,
this is equivalent to setting $u=0$ for all galaxies with radii $R>R_c$, so it is natural to interpret the cutoff radius $\Rc$ as a cluster radius.
In this light, it seems obvious that considerable care must be taken to choose the correct cluster radius when estimating
richness, but how to go about doing just that is a less straightforward question.

In this work, we propose that cluster radii be selected on the basis of a model radius-richness relation.  Specifically, we assume
that the size of a cluster of richness $\lambda$ scales as a power law of $\lambda$,
\begin{equation}
\Rc(\lambda) = R_0(\lambda/100.0)^\alpha.
\label{eq:radius}
\end{equation}
Naively, we expect $R_0\approx 1\ \Mpc$, as that is the characteristic size of clusters, and $\alpha\approx 1/3$ assuming
that $R\propto M^{1/3} \propto \lambda^{1/3}$.   We postpone the discussion of how we go about selecting $R_0$
and $\alpha$ to section \ref{sec:methods}.  For the time being, we shall simply assume that $R_0$ and $\alpha$
are known. In that case, equation \ref{eq:mfrichness} becomes
\begin{equation}
\lambda  = \sum p(\bx|\lambda) = \sum_{R<\Rc(\lambda)} \frac{\lambda u(\bx|\lambda)}{\lambda u(\bx|\lambda)+b(\bx)}.
\label{eq:richness1}
\end{equation}
Note that we have explicitly included the cutoff radius $\Rc$ in the sum above, and that this cutoff radius now depends
on $\lambda$.  Moreover, one can see that in the above equation, {\it the cluster richness $\lambda$ is the only 
unknown}, so we can numerically solve for $\lambda$.   In other words, by positing a richness-radius relation we are
able to simultaneously estimate both a cluster radius and the corresponding cluster richness.\footnote{Note 
that since we are explicitly setting $u=0$ for $R>R_c$, the fact that $u$ must be 
normalized to unity necessarily introduces a dependence of $u$ on $\lambda$.  That is, changing $\lambda$
will not only change the range of the sum in equation \ref{eq:richness1}, it will also change
the value of the summands.}


\subsection{The Filters}
\label{sec:filters}

In this work we consider three observable properties of galaxies:
$R$, the projected distance from a galaxy to the assigned cluster center,
$m$, the galaxy magnitude, and $c$, the galaxies' $g-r$ color.  We adopt
a separable filter function
\begin{equation}
u(\bx) = [2\pi R \Sigma(R)]\phi(m)G(c)
\end{equation}
where $\Sigma(R)$ is the two dimensional cluster galaxy density profile, $\phi(m)$ is the cluster luminosity function (expressed
in apparent magnitudes), and $G(c)$ 
is color distribution of cluster galaxies.  The prefactor $2\pi R$ in front of $\Sigma(R)$ accounts for the fact that given $\Sigma(R)$,
the radial probability density distribution is given by $2\pi R \Sigma(R)$.  Also, note
the separability condition makes the implicit assumption that these
three quantities are fully independent of each other, which is not true in 
detail \cite[for a discussion of the galaxy population of maxBCG clusters see][]{hansenetal07}.  
For instance, the tilt of the ridgeline implies that the mean color of a red sequence cluster galaxy varies slightly
as a function of magnitude.
We postpone an investigation of how including the correlation between these various observables affects our conclusions
to future work (Koester et al, in preparation).
We now describe each of our three filters in detail.  We note that defining said filters requires us to specify 
parameters governing the shape of the filters (e.g. $R_s$ for the radial filter, $\alpha$ for the luminosity
filter, etc.).  A detailed study on the dependence of our matched filter richness estimates on the shape of our filters
will be presented in future work.  

\subsubsection{The Radial Filter}

N-body simulations show that the matter distribution of massive halos can be well described by the 
so called NFW profile \citep[see e.g.][]{navarro_etal95,NFW},
\begin{equation}
\rho(r) \propto \frac{1}{(r/r_s)(1+r/r_s)^2}
\end{equation}
where $r_s$ is characteristic scale radius at which the logarithmic slope of the density profile is equal
to $-2$.  The corresponding two dimensional surface density profile \citep{bartelmann96} is 
\begin{equation}
\Sigma(R) \propto \frac{1}{(R/R_s)^2-1}f(R/R_s)
\end{equation}
where $R_s=r_s$ and
\begin{equation}
f(x) = 1-\frac{2}{\sqrt{x^2-1}}\tan^{-1}\sqrt{ \frac{x-1}{x+1} }.
\end{equation}
This formula assumes $x>1$. For $x<1$, one uses the identity $\tan^{-1}(ix) = i \tanh(x)$.  

Here, we assume that the NFW profile can also reasonably describe the density distribution
of galaxies in clusters \citep{linmohr04,hmwas05,popessoetal07}, and follow \citet{kmawe07b} in setting $R_s=150\ \kpc$.
In principle, one could optimize the value of this parameter, but we do not expect our final results
to be overly sensitive to our chosen value \citep[see e.g.][]{dongetal07}.
Also, in order to avoid the singularity at $R=0$ in the above expression, we set $\Sigma$ to a constant for 
$R\leq R_{core} =100\ \kpc$.
This core density is chosen so that the mass distribution $\Sigma(R)$ is continuous.  Our results are insensitive to
the particular choice of core radius for $R_{core} \leq 200\ \kpc$.
Finally, the profile $\Sigma(R)$ is truncated at the cluster radius $\Rc(\lambda)$, and is normalized
such that
\begin{equation}
1 = \int_0^{\Rc(\lambda)} dR\ 2\pi R\Sigma(R).
\end{equation}
We emphasize that this condition implies that the normalization constant for the density
profile is richness dependent, and must be recomputed for each $\lambda$ value when solving
for $\lambda$ in equation \ref{eq:richness1}.


\subsubsection{The Luminosity Filter}
At $z \lesssim 0.3$, the luminosity distribution of satellite cluster galaxies is well-represented by a 
Schechter function \citep[e.g][]{hansenetal07} which we write as
\begin{equation}
\phi(m) = 0.4 \ln(10)\phi_*10^{-0.4(m-m_*)(\alpha+1)}\exp\left(-10^{-0.4(m-m_*)}\right)
\end{equation}
We take $\alpha = 0.8$ independent of redshift.
The characteristic magnitude, $m_*$, is corrected
for the distance modulus, k-corrected, and passively-evolved using stellar population
synthesis models described in \citet{kmawe07a}. When applying the luminosity filter, 
$m_*$ is chosen from these models, appropriate to the
redshift of the cluster under consideration, and the filter is normalized by integrating down to a 
magnitude corresponding to $0.4L_*$ at the cluster redshift, or an absolute magnitude $M_i=-20.25$.  The latter is simply a luminosity cut
bright enough to make the maxBCG sample volume limited.


\subsubsection{The Color Filter}

Early type galaxies are known to dominate the inner regions of low redshift
galaxy clusters \citep[see e.g.][]{dressler84,kormendy89,hansenetal07}.  The
rest-frame spectra of these galaxies typically exhibit a significant drop at about 4000~\AA, that gives early type galaxies at the same redshift nearly uniformly red colors when observed through filters
that encompass this break.  In the SDSS survey, the corresponding filters for galaxies at $z\lesssim 0.35$ are $g$ and $r$, and 
we find that the $g-r$ colors of early type galaxies are found to be gaussianly distributed with a small intrinsic dispersion of about $0.05$ magnitudes.  Consequently, we take the color filter $G(c)$ to be
\begin{equation}
\label{eqn:color}
G(c|z) =\frac{1}{\sqrt{2\pi}\sigma}\exp \left[ \frac{(c-\avg{c|z})^2}{2\sigma^2} \right]
\end{equation}
where $c=g-r$ is the color of interest, $\avg{c|z}$ is the mean of the Gaussian color distribution of early type
galaxies at redshift $z$, and $\sigma$ is the width of the distribution.  The mean color $\avg{c|z}=0.625+3.149z$ was 
determined by matching maxBCG cluster members to the SDSS 
LRG \citep{eisensteinetal01} and MAIN \citep{straussetal02} spectroscopic galaxy samples.
The net dispersion $\sigma$ is taken to be the sum in quadrature of the intrinsic color dispersion $\sigma_{int}$,
set to $\sigma_{int}=0.05$,
and the estimated photometric error $\sigma_m$.  
In $g-r$, the typical photometric error on the red-sequence cluster galaxies brighter than $0.4L_*$ 
is $\sigma_m\approx 0.01 $ magnitudes for $z=0.1$, but can be as as large as
$\sigma_m\approx 0.05$ magnitudes for $z=0.3$.


\subsubsection{Background Estimation}

To fully specify our filters, we also need to describe our background model.  We assume the background
galaxy density is constant in space, so that $b(\bx)=2\pi R \bar \Sigma_g(m_i,c)$ where $\bar\Sigma_g(m_i,c)$
is the galaxy density as a function of galaxy $i-$band magnitude and $g-r$ color.   $\bar\Sigma_g(m_i,c)$
is estimated by distributing $10^6$ random points throughout the same SDSS photometric survey footprint that
defines our galaxy sample. All galaxies within an angular separation of $0.05$ degrees of the random points
(about $1\hMpc$ at $z=0.25$) are used to empirically determine the mean galaxy density $\bar \Sigma_g(m_i,c)$
using a top hat  cloud-in-cells (CIC) algorithm \citep[e.g.][]{hockney81}.  For our cells, we used
60 evenly-spaced bins 
in $g-r \in [0,2]$ and 40 bins in $i \in [14,20]$. In each 2 dimensional bin, the number density of galaxies
is normalized by the total number of random points, the width of each color and magnitude 
bin (0.05 mags and 0.1 mags, respectively), and area searched ($0.05^2 \pi$ degrees). 

This process creates an estimate of the global background, i.e. the number density of galaxies as a function 
of color and magnitude in the full SDSS survey. Not surprisingly, a similar result is obtained by binning
the whole galaxy catalog in color and magnitude with CIC and dividing by the survey area. However, the 
procedure we employ above can readily be adapted to returning alternative background estimates, e.g the local 
cluster density as a function of redshift, by replacing random points with clusters.


\section{Methods}
\label{sec:methods}

We have now fully specified our richness estimators, except for the values
$R_0$ and $\alpha$ that govern the radius-richness scaling relation.  We now
discuss how we go about selecting optimal values for these parameters.

As we mentioned earlier, we wish to find the cluster richness estimator that
minimizes the scatter in the richness-mass relation.  Cluster mass, however, is
not an observable, and thus we must rely on other mass tracers.  Here, we 
use X-ray luminosity ($L_X$) as our mass proxy, primarily because it is a well
known mass tracer~\cite[e.g.][]{rb02,sebsn06,rembj08} that is readily
accessible to us and for which we can quickly estimate the scatter for multiple
richness measures \citep[see][]{rmbej08}. 

We proceed as follows: we begin by defining a coarse grid in $R_0$ and $\alpha$,
given by
\begin{eqnarray}
R_0 & = & \{0.5,0.75,1.0,1.25,1.5\}\, \\
\alpha & = & \{-0.05,0.05,0.15,0.25,0.35,0.45\}
\end{eqnarray}
where $R_0$ is measured in units of $\hMpc$.
Each of these grid points defines a distinct richness estimator through
equation \ref{eq:richness1}.  For each grid point, we estimate the
corresponding richness for every cluster in the maxBCG catalog.  We then select
the 2000 richest clusters and calculate the scatter in $L_X$ at fixed richness
of those top 2000 clusters.  Note that, because the rank ordering of the clusters changes
as we vary our richness estimate, the clusters used to estimate the scatter in $\Lx$
varies slightly across the grid.  We limit ourselves to the richest 2000 clusters to
ensure our results are insensitive to the $N_{200}\geq 10$ cut in the maxBCG catalog. 

From our measurements of the scatter
$\sigma_{\lnl|\lambda}(R_0,\alpha)$ at each grid point, we can
directly read which parameter combination minimizes the scatter.  We emphasize
that because the scatter in mass at fixed $\Lx$ is much lower than the
corresponding scatter at fixed richness~\citep{rozoetal08a}, for our
purposes $\Lx$ is a nearly perfect mass tracer.  We note that the X-ray
measurements described in \ref{sec:xray} require a minimum aperture of
$500\,\kpc$.  For the 2000 richest clusters, this cutoff is only employed when
$R_0 = 0.5\,\hMpc$ and $\alpha \ge 0.15$, which is a region of parameter space
that already does not appear to have a strong correlation between $L_X$ and
richness.  Therefore, we conclude that the aperture cutoff does not have a
significant effect on our results.

To determine the uncertainty in the recovered parameters $R_0$ and $\alpha$, we need to understand the errors 
in our measurement of the Lx-richness scatter. We estimate these errors using bootstrap resampling.
We proceed as follows:
let $\mu$ be an index that runs over all grid points $(R_0,\alpha)$, and $\sigma_\mu$
be the scatter at the $\mu^{th}$ grid point.
We resample (with replacement) the full maxBCG catalog, and measure the scatter 
$\sigma_\mu$ at every grid point.  The procedure is iterated 100 times, and the
measurements are used to estimate the mean and covariance matrix 
of $\sigma_\mu$.\footnote{The
measurement of the scatter in $\Lx$ at fixed richness is very time consuming,
and needs to be done independently for every point in the grid.  This explains
why we restrict ourselves to only 100 bootstrap resamplings.}  Assuming
that the probability distribution $P(\sigma_\mu)$ is a multi-variate Gaussian characterized
by the observed mean and covariance matrix, we generate
$10^5$ Monte Carlo realizations of the scatter, and estimate the
fraction of times that each grid point is observed to have the lowest scatter among all
grid points.

To use the grid to zero in on a particular value for $R_0$ and $\alpha$, and to
estimate errors in these values, we fit each of the $10^5$
realizations of the scatter $\sigma_{\lnl|\lambda}(R_0,\alpha)$ with a 2D parabola.  From the
fits, we can read off the values of $R_0$ and $\alpha$ at which the minimum 
occurs, giving us $10^5$ samplings of the probability distribution of the
location of the minimum in parameter space.  The probability distribution of the
resulting $10^5$ minima is exactly what we desired.

As it turns out, and as discussed in \S~\ref{sec:results}, the coarse grid
defined above is too broad for a parabolic fit to adequately describe 
the function $\sigma_{\lnl|\lambda}(R_0,\alpha)$.  However, if we restrict
ourselves to a smaller region of parameter space near the minimum
determined from the coarse grid, a quadratic fit becomes adequate. Therefore, we have defined a narrower fine grid,
\begin{eqnarray}
R_0 & = & \{1.0,1.1,1.2,1.3,1.4\}\\
\alpha & = & \{0.22,0.26,0.30,0.34,0.38,0.42\}
\end{eqnarray}  
with $R_0$ measured in units of $\hMpc$.
It is this grid that we use to report our final results and to select the
optimal parameters $R_0$ and $\alpha$.

To summarize, we first do a rough exploration of the parameter space 
$R_0$ and $\alpha$ using a coarse grid, and then use a smaller but
finer grid to statistically constrain the location of the scatter minimum.


\section{Results}
\label{sec:results}

\subsection{The Full Sample}

\begin{figure}
\begin{center}
\scalebox{0.7}{\rotatebox{270}{\plotone{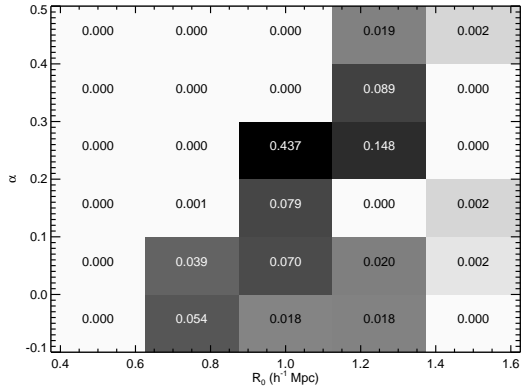}}}
\caption{Probability that a given point in the grid minimizes the scatter in $\Lx$ at fixed
richness in the coarse grid.  The gray scale varies logarithmically with the probability,
which is explicitly quoted in the Figure.  Note the broad
  degeneracy region from $(R_0,\alpha)\approx (0.8,0.0)$ to
  $(R_0,\alpha)\approx (1.4,0.5)$, where the scatter $\sigma_{\lnl|\lambda}
  \sim 0.78$.  }
\label{fig:coarsegray}
\end{center}
\end{figure}

Figure~\ref{fig:coarsegray} illustrates the probability that each coarse grid
point is found to minimize the scatter of the 2000 richest clusters when
resampling our data as described in \S~\ref{sec:methods}.  For this plot, we
have used the full cluster sample, though a similar result holds when using the
clean cluster sample.  Each square is shaded in gray on a log scale according
to the fraction of trials that point is found to have the minimum scatter.  The
primary feature of this plot is a broad degeneracy region from
$(R_0,\alpha)\approx (0.8,0.0)$ to $(R_0,\alpha)\approx (1.4,0.5)$,
corresponding to a scatter $\sigma_{\lnl|\lambda} \approx 0.78$.  Note
this scatter is a significant improvement relative to the $\Lx-$richness
scatter measured for $\Nobs$, $\sigma_{\lnl|N_{200}}=0.96$.
The scatter in $\Lx$ increases as we move away from the degeneracy
region, ranging from $\sigma_{\lnl|\lambda} \sim 0.86$ in the lower-right corner 
of Figure \ref{fig:coarsegray} to
$\sigma_{\lnl|\lambda} > 1.0$ in the upper-left corner.  Further discussion
of why our new richness estimator results in significantly reduced scatter
is presented in \S~\ref{sec:zdep}.


\begin{figure}
\begin{center}
\scalebox{0.7}{\rotatebox{270}{\plotone{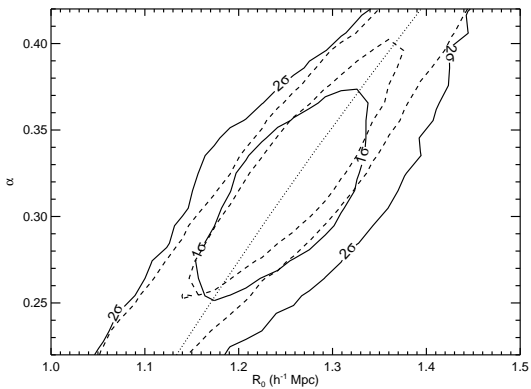}}}
\caption{\label{fig:finecombo}Contour plot of the probability density of the
  points in $R_0-\alpha$ space that minimize the scatter $\sigma_{\lnl|\lambda}(R_0,\alpha)$.
  The solid contours show the $1\sigma$ and $2\sigma$
  contours for the full sample, and the dashed lines show the same contours for
  the ``cleaned'' sample (see \S~\ref{sec:cleaning}).  The closing of the
  $1\sigma$ contours in the upper-right and lower-left are likely an
  artifact of the grid.  The dotted line shows the contour of fixed mean
  cluster radius $R_c=900\,\kpc$.  All the richness estimators along this line result
  in the same mean cluster radius, and have therefore very similar richness values.}
\end{center}
\end{figure}

Figure~\ref{fig:finecombo} shows the probability density of the points in
$R_0-\alpha$ space that minimize the scatter in $\Lx$ at fixed richness for the
fine grid, as estimated through the parabolic fits to the function
$\sigma_{\lnl|\lambda}(R_0,\alpha)$ described in section \S~\ref{sec:methods}.
The solid contours are for the full cluster sample and the dashed contours are
for the clean cluster sample.  The diagonal degeneracy suggested in the
previous plot is now very obvious, especially in the $2\sigma$
contour. Importantly, both the full and clean sample produce very similar
results, although the contours are noticeably smoother for the clean sample.
We note that the closing of the $1\sigma$ contours in the upper-right and
lower-left is likely an artifact of the grid boundaries.  As demonstrated in
the coarse grid in Figure~\ref{fig:coarsegray}, the degeneracy region extends
at least to $\alpha\sim0$ and $\alpha\sim0.5$.

The existence of the degeneracy region is relatively simple to explain.
Consider the problem we are trying to address: what is the correct size of a
cluster?  Roughly speaking, this involves two parts: one, determining the
correct cluster size of the average cluster, and two, determining how the
cluster size scales with richness as one moves away from the average cluster.
The former is much better determined than the latter, so in the $(R_0,\alpha)$
plane, one typically expects a sharp constraint on the mean cluster radius, and a
considerably weaker constraint on the orthogonal direction, corresponding to
the scaling of the radius with richness around the statistical pivot point.
Thus, we expect the observed degeneracy between $R_0$ and $\alpha$ to pick out
parameter combinations that hold the median cluster radius of the sample fixed.

Figure \ref{fig:finecombo} clearly illustrates that this is the case.
In the figure, the diagonal dotted line corresponds to a contour of fixed
median cluster radius $\tilde R(R_0,\alpha)=900\,\kpc$, where the function 
$\tilde R(R_0,\alpha)$ is defined as the median cluster radius of the 2000
richest clusters.   The fact that this contour falls almost exactly along the observed
degeneracy between $R_0$ and $\alpha$ strongly supports our interpretation.


Our argument suggests a way to break the degeneracy between $R_0$ and
$\alpha$.  If we can measure the scatter in $\Lx$ at fixed richness at two
very different richness scales, then the mean radius picked out by each
of the samples will be substantially different.   This, in turn, rotates the
degeneracy lines relative to each other, so that the intersection defined
by the two samples would cleanly pick out a single value for $R_0$
and $\alpha$.

We have repeated our analysis on the top 500 and 1000 clusters, but these
thresholds are much too close to our reported 2000 clusters to be able to
successfully break the observed degeneracy.  Ideally, we would repeat our study
using the 10000 or 20000 richest clusters, thereby guaranteeing a degeneracy
region that is significantly rotated relative to that of Figure
\ref{fig:finecombo}.  Unfortunately, performing our scatter analysis on the top
10000 clusters is not presently possible since the vast majority of this larger
cluster sample does not emit sufficiently in X-rays to allow for individual
luminosity estimates of the clusters.  Furthermore, when choosing more than the
top $\sim3000$ clusters we begin to run into threshold effects due to the
initial selection of maxBCG clusters with $N_{200}\ge 10$.  One might hope
instead to repeat our analysis using not the top 10000 clusters, but rather the
top 100 clusters, that is, by limiting ourselves to the very richest systems.
Unfortunately, this suffers from a different problem: when looking at the top
100 clusters only, the range of richnesses being sampled is much too narrow to
allow a simultaneous estimate of the amplitude, slope, and scatter of the
$L_X$-richness relation, so performing our analysis using the top 100 clusters
only is also not feasible. Thus, at the time being, we must simply accept the
existence of a large degeneracy between $R_0$ and $\alpha$.


\subsection{Selecting an Optimal $\alpha$}

Due to the large degeneracy between $R_0$ and $\alpha$, it is difficult to
select any single point in $R_0-\alpha$ space as optimal.  We note, however,
that the degeneracy region goes through $\alpha=1/3$, which is loosely
theoretically motivated based on the naive expectation $R^3\propto M \propto
\lambda$.  Since our goal is to define a unique richness measure, we have opted
for setting $\alpha=1/3$.  Given that the degeneracy region goes through
$\alpha=1/3$, our choice does not adversely affect the properties of our
richness estimator.  That is, the scatter for $\alpha=1/3$ is indistinguishable
from that of the best possible value for $\alpha$ to within observational
uncertainties.

Using a principal component analysis on the best-fit minima that describe the
contours in Figure~\ref{fig:finecombo}, we have calculated the 
degeneracy axis for each of the full and clean cluster samples.  
For the full cluster sample we obtain
\begin{equation}
\ln (R_0/1\ h^{-1}\ \Mpc) - 1.342(\alpha-0.33) = 0.25 \pm 0.04,
\end{equation}
while for the clean cluster sample we find
\begin{equation}
\ln (R_0/1\ h^{-1}\ \Mpc) - 1.277(\alpha-0.33) = 0.24 \pm 0.03
\end{equation}
We have confirmed that the residuals are Gaussian along most of the degeneracy
axis.  We quote the degeneracy line in terms of $\ln R_0$ and $\alpha$ rather than
$R_0$ and $\alpha$ themselves simply because the former results in more accurate
extrapolations for $\alpha$ values that are very different form $\alpha=1/3$.

We are
encouraged by the fact that the clean and full samples give fully consistent
results, thus showing that the known cool core clusters and obvious foreground
contamination are not significantly biasing the best combination of $R_0$ and
$\alpha$.  Our final choice for $R_0$ and $\alpha$ is therefore
$R_0=1.27\ h^{-1}\ \Mpc$ and $\alpha=1/3$.


\subsection{Improvement in the Scatter}

\begin{figure}
\begin{center}
\scalebox{1.2}{\rotatebox{0}{\plotone{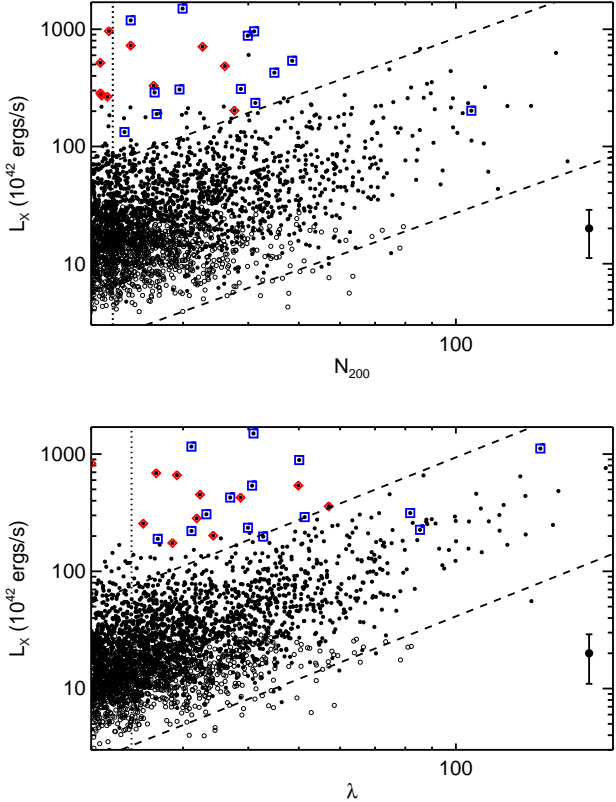}}}
\caption{\emph{Top panel:} $L_X$ vs. $N_{200}$ for the 3000 richest clusters.
  Following \citet{rmbej08}, the solid points represent $>1\sigma$ detections,
  and the empty circles represent $1\sigma$ upper limits.  The vertical dotted
  line represents the cutoff for the top 2000 clusters used in the analysis.
  The dashed lines represent the $\pm2\sigma_{\ln L|N_{200}}$ scatter
  constraints. The fictitious data point in the lower-right corner shows the
  typical $\Lx$ error.  The red diamonds represent clusters that are excluded
  from the clean sample because they are obviously contaminated by foreground
  X-ray emission.  The blue squares represent clusters that are excluded from
  the clean sample because they are known cool core clusters.
  \emph{Bottom panel:} $L_X$ vs. $\lambda$ for $R_0 =
  1.27$, $\alpha=1/3$ for the 3000 richest clusters; the symbols are the same
  as for the top panel. Our optimized matched filter richness estimate $\lambda$ is significantly more tightly
  correlated with $\Lx$ than $N_{200}$.}
\label{fig:lxvslam}
\end{center}
\end{figure}

Now that we have a fully specified $R_0 = 1.27\ \hMpc$ and $\alpha=1/3$, we
have measured the matched filter richness of every cluster in the
\citet{kmawe07a} sample. Figure~\ref{fig:lxvslam} shows $L_X$ vs. $N_{200}$
(top panel) and $L_X$ vs. $\lambda$ (bottom panel) for the top 3000 richest
clusters.  Following \citet{rmbej08}, the solid points represent detections at
the $>1\sigma$ level, and the empty points represent $1\sigma$ upper limits.
The vertical dotted line represents the cutoff for the top 2000 richest
clusters used in this analysis.  Though not obviously visible in this plot, the scatter in
$\lambda$ is significantly decreased.  We note that there are still some
significant outliers in the $\Lx-\lambda$ relation, especially at high $\Lx$.
The red diamonds and blue squares represent clusters that are removed from the
clean cluster sample.  The red diamonds are clusters whose measured X-ray flux
is known to be contaminated by foreground emission from stars, nearby galaxy
clusters, or AGN.  The blue squares represent the known cool core clusters.
These are, for the most part, significantly brighter than typical maxBCG
clusters at similar richness, which is consistent with the hypothesis that the
X-ray luminosity of these clusters is boosted by emission from the core.

Table~\ref{tab:sigmas} summarizes how the scatter of the 2000 richest clusters
varies as we change our richness measure.  Here, we consider three richness
measures only: $N_{200}$, which is the original richness estimate for maxBCG
clusters presented in \citet{kmawe07b}; $N_{200}L_{BCG}^{0.79}$, which was
suggested by \citet{reyesetal08} as an improvement over $N_{200}$ by making use
of $L_{BCG}$, the luminosity of the cluster BCG; and our optimized matched
filter richness estimator $\lambda$.  We see that for both the full and clean
sample, our optimized matched filter estimator significantly outperforms both
$N_{200}$ and $N_{200}L_{BCG}^{0.79}$.  To quantify the significance of the
improvement, we must take into account the fact that the errors are correlated.
Following \S~\ref{sec:methods}, we have performed bootstrap resampling on the
full catalog and clean catalog, calculating the scatter in the top 2000
clusters for both $\lambda$ and $N_{200}$.  For each bootstrap resampling we
calculate $r=\sigma_{\lnl|\lambda}/\sigma_{\lnl|N_{200}}$.  The deviation from
$r=1.0$ can be used to quantify the significance of the improvement.  The
improvement in the scatter relative to $\Nobs$ is significant at $9\sigma$ for
the full cluster sample, and at $11\sigma$ for the clean sample.

\begin{deluxetable}{ccc}
\tablewidth{0pt}
\tablecaption{\label{tab:sigmas}Scatter in $L_X$ at fixed richness, top 2000
  clusters}
\tablehead{
\colhead{Richness} & \colhead{Full Sample} & \colhead{Clean Sample}
}
\startdata
$N_{200}$ & $0.95\pm0.03$& $0.86\pm0.02$\\
$N_{200}L_{BCG}^{0.79}$ & $0.84\pm0.02$ & $0.78\pm0.02$\\
$\lambda$ & $0.79\pm0.02$ & $0.70\pm0.02$\\
$\lambda$ & $0.78\pm0.02$ & $0.69\pm0.02$ \\
\enddata
\tablenotetext{}{Except for the last row, $\Lx$ was measured within a fixed $1\ \hMpc$ aperture.
The scatter in $\Lx$ quoted in the last row is different only in that it measured $\Lx$ within the assigned optical  cluster radius 
$R_c(\lambda)$.
The combination $\Nobs L_{BCG}^{0.79}$ was suggested by \citet{reyesetal08} as an improvement
over $\Nobs$.  The error bars define $68\%$ confidence intervals.}
\end{deluxetable}


\section{Redshift Dependence}
\label{sec:zdep}

\citet{rmbej08} showed that there is strong redshift evolution in the
$\avg{\Lx|\Nobs}$ relation of maxBCG clusters.  Similar redshift dependence
is observed in the velocity dispersion-optical richness relation measured in
\citet{beckeretal07}.  This is best understood as a variation of $\Nobs$ at
fixed mass, with an observed fractional decrease in $\Nobs$ of $30\%-40\%$
over the redshift range of the maxBCG catalog.  In our previous work, the
origin of this redshift dependence was unclear.  Here, we demonstrate how the
matched-filter richness removes this redshift dependence, and show the pitfalls
of a simple richness estimator such as $\Nobs$.

\begin{figure}
\begin{center}
\scalebox{0.7}{\rotatebox{270}{\plotone{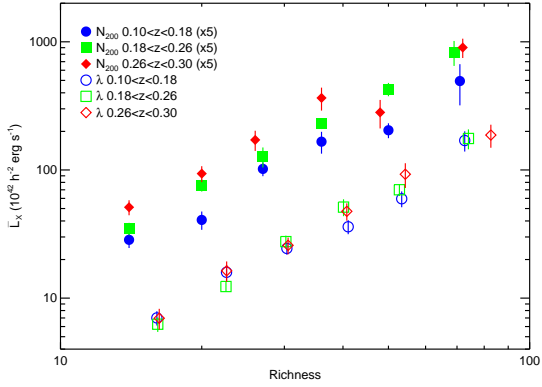}}}
\caption{$\avg{\Lx}$ vs. richness in three different richness bins.  The empty
  points denote the matched filter richness $\lambda$, and the solid points
  denote the original maxBCG richness $N_{200}$.  The three richness bins are:
  $0.10<z<0.18$ (blue circles); $0.18<z<0.26$ (green squares); $0.26<z<0.30$
  (red diamonds).  The normalization of $\avg{\Lx}-N_{200}$ has been multiplied
  by 5 for clarity.  It is readily apparent that $N_{200}$ has a strong
  redshift dependence~\citep{rmbej08, beckeretal07}, while $\lambda$ does not.
  }
\label{fig:evol}
\end{center}
\end{figure}

Figure~\ref{fig:evol} shows the $\avg{\Lx|\Nobs}$ relation for maxBCG
clusters split into three different redshift bins (solid symbols).  Also shown
is the mean relation $\avg{\Lx|\lambda}$ for the same three redshift bins
(empty symbols).  It is obvious from the figure that the redshift evolution in
the $\Lx-$richness relation is significantly weaker for $\lambda$ than it is
for $\Nobs$.  We have fit the data with a power-law evolution in redshift,
following \citet[][\S 5.3]{rmbej08}:
\begin{equation}
\avg{\Lx|N} = A \left( \frac{N}{40} \right)^\alpha \left( \frac{1+z}{1+\tilde z} \right)^\gamma
\end{equation}
where $\tilde z$ is the median redshift of the cluster sample and $N$ is the
richness measure of interest.  We find that $\gamma=6.0\pm0.8$ for $\Nobs$
while $\gamma=0.7\pm0.8$ for $\lambda$, consistent with no evolution.

Note, however, that even if the relation between $\lambda$ and cluster mass is
redshift independent, we expect to see evolution in the $\Lx-\lambda$ relation
due to evolution in the $\Lx-M$ relation.
The expectation for self-similar evolution in $\Lx$ at fixed mass is
that $\Lx \propto \rho_c(z)^{7/6}$ for bolometric luminosities, but closer to
$\bar \rho_c^{1.0}$ for soft-band X-ray luminosities \citep{k86}.  Here, $\rho_c$
is the critical density of the universe at redshift $z$.
In a ${\Lambda}$CDM universe with
$\Omega_m = 0.25$, the expected soft X-ray band evolution is thus $\gamma \approx 1.05$,
so our results are also consistent with self-similar evolution.

The striking difference in the evolution in the $\Lx-$richness relation between
$\lambda$ and $N_{200}$ is due to the differences in how $\Nobs$ and $\lambda$
employ galaxy colors when estimating cluster richness.
For $\Nobs$, a galaxy contributes to the richness
if and only if its color differs from the BCG color by no more than
twice the intrinsic width of the ridgeline color width plus the galaxy's
photometric error, added in quadrature.  That is, 
$\Nobs$ weighs galaxies according to the probability distribution $p_{top-hat}(c)$
given by:
\begin{equation}
p_{top-hat}(c) = \left \{ \begin{array}{cl} 
	1 & \mbox{if}\  |c-c_{BCG}| \leq \sqrt{(2\sigma_{int})^2+\sigma_{obs}^2} \\
	0 & \mbox{otherwise}
	\end{array} \right .
\end{equation}
where $\sigma_{int}=0.05$ is the intrinsic width of the ridgeline.
This is a top-hat distribution in observed color, but the width of the top-hat
depends on the photometric error of the galaxy under consideration. 
Also, note that the center of the color box is not the model $\avg{c|z}$
quoted earlier, but rather the color of the BCG, which, as we show below,
is a very significant difference.

To illustrate how these differences in the color filter results in differences
in the evolution and scatter of $\lambda$ and $\Nobs$, we have defined three
additional richness measures with key properties bridging those of $\lambda$
and $\Nobs$.  Including $\lambda$ and $\Nobs$, the five richness measures
considered here are
\begin{enumerate}
\item{$\lambda$: the matched filter richness with a variable aperture, as
  described above, with a gaussian color filter centered on $\avg{c|z}$.}
\item{$\lambda_{BCG}$: the matched filter richness using the same aperture as
  with $\lambda$, but with the Gaussian model centered on $c_{BCG}$.}
\item{$N_{top-hat,model}$, a top-hat richness using the $p_{top-hat}$ formulation
  above, centered around $\avg{c|z}$ as in Eqn.~\ref{eqn:color}, measured on a
  fixed $1\,\hMpc$ scale.}
\item{$N_{top-hat,BCG}$, a top-hat richness using the $p_{top-hat}$ formulation
  above, centered around $c_{BCG}$, measured on a fixed $1\,\hMpc$ scale.  This
  is similar to the maxBCG $N_{gals}$ richness, without the additional cut on
  the $r-i$ color of the member galaxies.}
\item{$\Nobs$, the original maxBCG richness estimator, measured in a scaled
  radius $r_{200}^{gals}$, with the color filter centered on $c_{BCG}$.}
\end{enumerate}

Table~\ref{tab:evol} shows the scatter (in the top 2000 clusters) and evolution
parameters for these various richness estimators.  There are two key
observations that we can make here.  First, when using the top-hat richness,
centering around the model color is significantly better than centering on the
BCG color, in terms of decreasing both the scatter and evolution of the
richness measure.  Second, the smooth Gaussian filter centered on the BCG color
works almost as well as the Gaussian filter centered on the model color.  This
is a significant result, because it implies that not only are the resulting
richnesses more robust to moderate changes in the color filter parameters, but
also the richness measure itself is also robust to photometric redshift errors.
The reason for this robustness is simple: when using a color top-hat selection,
using the correct color model is of paramount of importance since miscentering
of the top-hat will lead to underestimates of the richness.  In the matched
filter framework, what is important is the relative galaxy density of the
cluster and field components, which can remain high even if the centering of
the ridgeline color is slightly displaced.  Thus, matched-filter richness
estimates are much more robust to small changes in the parameters of the color
filter than estimates based on simple color cuts.

\begin{figure}
\begin{center}
\scalebox{0.7}{\rotatebox{270}{\plotone{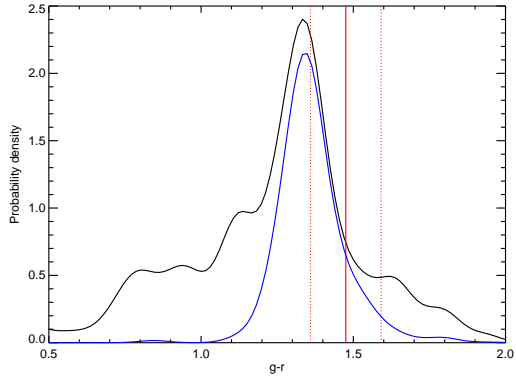}}}
\caption{\label{fig:clust983colors}Color distribution of all the galaxies
  brighter than $0.4\,L_*$ within $1\,\hMpc$ (solid black line) of the galaxy
  cluster SDSS~J082026.8+073650.1 at a redshift of $z_{spec}=0.22$.  
  The distribution is estimated using a Gaussian Kernel Density Estimator (KDE), with the
  size of the kernel selected to adequately sample the peak due to ridgeline galaxies.  
  The cluster BCG color (solid red line) is significantly redder than the red
  sequence (peak of the black distribution).  The dotted vertical lines show the
  $\pm2\sigma_{int}$ color cut, which does not include the bulk of the red
  sequence, and therefore $N_{200}$ is significantly underestimated.  The
  blue curve  is the KDE estimate of the galaxy distribution,  except  every galaxy 
  has been weighted by its membership probability as estimated
  using the matched filter approach.  We can see the match filter richness estimate
  selects principally ridgeline galaxies.
}
\end{center}
\end{figure}

As an illustration of this effect, Figure~\ref{fig:clust983colors} shows the color distribution
of all galaxies brighter than $0.4\,L_*$ within
$1\,\hMpc$ (solid black line) of the galaxy cluster SDSS~J082026.8+073650.1 at
a redshift $z_{spec}=0.22$.  This cluster was selected because of the large
discrepancy between $N_{200}$ and $\lambda$.  The color of the cluster BCG
(solid red line) is significantly redder than the red sequence.  The dotted
vertical lines show the $\pm2\sigma_{int}$ color cut, which does not include
the peak of the red sequence.  As a result, $N_{200}$ is significantly
underestimated in this system.   The blue curve shows the same galaxy distribution,
but weighing each galaxy by its membership probability as estimated using the
matched filter approach.
As we can see, the matched filter effectively selects galaxies in the red sequence.

\begin{deluxetable}{ccc}
\tablewidth{0pt}
\tablecaption{\label{tab:evol}Scatter ($\sigma_{\lnl|N}$) and redshift
  evolution ($\gamma$)}
\tablehead{
\colhead{Richness} & \colhead{$\sigma_{\lnl|N}$}\tablenotemark{a} &
\colhead{$\gamma$}
}
\startdata
$\lambda$ & $0.78\pm0.02$ & $0.7\pm0.8$\\
$\lambda_{BCG}$ & $0.82\pm0.02$ & $1.1\pm0.8$ \\
$N_{top-hat,model}$ & $0.80\pm0.02$  & $0.5\pm0.8$\\
$N_{top-hat,BCG}$ & $0.99\pm0.02$ & $4.2\pm0.7$\\
$\Nobs$ & $0.95\pm0.02$ & $6.0\pm0.8$\\
\enddata
\tablenotetext{a}{For the top 2000 clusters}
\end{deluxetable}

We have demonstrated that the redshift evolution observed in the $\Lx-\Nobs$
relation is primarily caused by using a top-hat filter centered on the color of the BCG.
Why such a choice of color filter results in the strong evolution we observe for 
$N_{200}$ is a complicated question, with at least two physical mechanisms
contributing to the problem at comparable levels.  First, there is the fact that
even for a correctly centered top-hat filter, a ridgeline galaxy can fall outside
the color cuts due to photometric errors.  Since photometric errors increase
with increasing redshift, a color cut such as that of $\Nobs$ will progressively
lose more galaxies as one moves the cluster to higher redshift.
Second, the E/S0 ridgeline is not flat, but
has a slight tilt ($\sim -0.04$ mags/mag in $g-r$ vs. $i$), such that brighter
galaxies tend to be redder \citep[e.g.][]{visvanathan77, renzini06}.  By
centering the color filter on the BCG -- by definition the brightest and
usually reddest cluster member -- a small richness bias is introduced: clusters
with brighter BCGs have a color filter centered redward of the average BCG
color.  Moreover, recent work by
Hao et al. (in preparation) shows that with a proper account for photometric
errors, the ridgeline tilt evolves with redshift, such that the ridgeline is
\emph{steeper} at $z=0.3$ than at $z=0.1$.  Consequently, a BCG centered color
cut becomes increasingly offset from the true mean ridgeline color as we increase
redshift.  Both of these systematics effects occur with similar magnitude, and act
in concert to produce the observed evolution in $\Nobs$.   We emphasize,
however, that our matched filter richness estimator does {\it not} suffer from
these systematic effects.

Finally, we can now also explain why $N_{200}$ exhibits stronger evolution than
$N_{top-hat,BCG}$.  Recall that the aperture used to estimate $\Nobs$ is itself
based on the richness measure $N_{gals}$, which is very similar to
$N_{top-hat,BCG}$.  Since $N_{top-hat,BCG}$ systematically underestimates the
richness for high redshift clusters due to the increasing tilt of the
ridgeline, the aperture $r_{200}^{gals}$, which scales with $N_{top-hat,BCG}$,
is also underestimated.  This compounds the effect of incorrect centering of the color
box and results in stronger redshift evolution.


\section{Summary and Conclusions}
\label{sec:conclusions}

We have introduced a new matched filter richness estimator $\lambda$ whose correlation with
mass is significantly tighter than that of $\Nobs$, the original maxBCG richness estimator.
Relative to other matched filter estimates, our estimator has two significant differences:
\begin{enumerate}
\item The richness is measured on a scale that is optimized in the sense that it minimizes the scatter in $\Lx$ at
fixed richness.  
\item In addition to a radial and magnitude filters, we include a color filter.  This is of crucial importance
for differentiating between member and non-member (projected) galaxies.

\end{enumerate}
The first these points is important since we have demonstrated
that a poor choice of aperture increases the scatter in mass at
fixed richness, while the latter minimizes the impact of
projection effects in richness estimates.  Of the two, however, the improved treatment of
galaxy color is the principal reason for the marked reduction of the scatter in the
$\Lx-$richness relation.

Our procedure for aperture optimization can be easily generalized to any mass tracer for
which one can construct a calibrating data set.  In our particular case, we minimize the
scatter in the $\Lx-\lambda$ relation by measuring both $\Lx$ and $\lambda$ within an aperture 
$R_c(\lambda)=R_0(\lambda/100)^\alpha$, and varying the model parameters $R_0$
and $\lambda$.  Given the small richness range probed by our sample,
we have not been able to isolate unique values for $R_0$ and $\alpha$, finding instead
a degeneracy region corresponding to a fixed mean cluster radius for the clusters in
the sample.   Based on \emph{a priori} assumptions about the
radius--richness scaling, we have fixed $\alpha=1/3$, which yields a
normalization of $R_0 = 1.27\pm0.03$.  We note, however, that
the degeneracy region intersects $\alpha=0$ at $R_0\approx 850\ \kpc$.
Although we expect that this fixed scaling will not be
ideal at the rich group/poor cluster scale, it does work as a ``first guess''
richness and may be applicable to future cluster finding techniques.
At this point, it is unclear whether the cluster radii selected by our technique
reflects a true physical property of the maxBCG clusters, or whether it is driven
primarily by a compromise between the the increase signal one expects at larger
aperture, and the smaller noise one expects for smaller apertures.
Regardless of the source, it is likely that similar aperture dependences exist for other 
mass tracers.  

We have also found our new richness estimator has scaling relations whose
redshift evolution is much more mild than those exhibited by $\Nobs$.  This
difference arises due to two effects: first, $\Nobs$ uses a top-hat filter to
select cluster galaxies, where as our matched filter estimator $\lambda$ uses
gaussian color filters.  Second, $\Nobs$ centered its color filter at the color of
the BCG, whereas $\lambda$ centered its color using an observationally
calibrated color--redshift relation.  The fact that the color of the BCG does
not always agree with the observationally calibrated redshift-color relation
leads to a systematic difference between the two richness measures.  Moreover,
we also found that the sharp edges of the top-hat filter result in a richness
estimator that is very sensitive to the details of the color model, whereas our
gaussian filter is much more robust to moderate changes in the model parameters.

Restricting ourselves to the clean cluster sample, which excludes cooling flow clusters and
clusters with obvious foreground contamination in their X-ray luminosities, we have found that
the scatter in the $\Lx-$richness relation of the 2000 richest clusters is
is $\sigma_{\lnl|\lambda}=0.69$ for $\lambda$, compared to $\sigma_{\lnl|\Nobs}=0.86$ for $\Nobs$.  
Assuming a slope of $\approx 1.6$ for the $\Lx-M$ relation~\citep{sebsn06,rembj08,vbefh08}, 
these amount to a logarithmic scatter in the mass--richness relation of $\approx 0.43$ and $0.54$ respectively.
While this is a very significant improvement, we expect that further tightening
of the scatter in mass at fixed richness must be possible.
For instance, assuming the intrinsic scatter in the richness-mass relation
is Poisson, the logarithmic scatter possible for clusters with 20 galaxies or so should be 
roughly $\approx 0.2$.  

Fortunately, there
are still many options left for us to explore in our quest to define optical mass proxies that can
be competitive with other mass tracers in terms of the tightness of the correlation with mass. As we have
defined it here, our richness estimates only makes use of the number of galaxies in the cluster.
One could, for instance, weigh our cluster richness by other optical mass tracers such as
the luminosity of 
the brightest cluster galaxy \citep{reyesetal08}, the abundance of baryons 
contained in the intracluster light \citep[e.g.][]{gonzales07}, or other aspects of the cluster galaxy
morphology \citep[e.g. Bautz-Morgan Type, ][]{bautz70}.  In addition, one could weigh each galaxy's contribution
to the richness by physical observables such as galaxy luminosity.  Such a luminosity weighted richness
estimate would be a measure of the optical luminosity of the cluster as a whole, and might be 
better correlated with mass than richness itself \citep[see
also][]{lin03,milleretal05, pbbrv05}.
It is also likely that
further improvements in richness estimates can arise with more accurate filters, a possibility we intend
to explore in future work.  Finally, we know that even with today's filters, part of the scatter we observe 
must be due to systematics effects such as failures of the cluster finding algorithm in identifying the correct
center of a cluster, a problem which we have not addressed in this work.
For the time being,
the fact that naive theoretical expectations result in a scatter much lower than previously observed,
and the fact that on our first attempt at defining a better richness estimator resulted in a highly
significant ($\approx 11\sigma$) improvement over $\Nobs$, suggest that the future is rife with opportunities
for this kind of work.

\acknowledgements

 ER would like to thank David Weinberg and Christopher Kochanek for interesting
discussions and their careful reading of the manuscript. ESR would like to thank the TABASGO foundation.
RHW was supported in part by the U.S. Department of Energy under contract
number DE-AC02-76SF00515 and by a Terman Fellowship at Stanford
University.  TM and JH gratefully acknowledge support from NSF grant AST 0807304 and DoE Grant DE-FG02-95ER40899.
This project was made possible by workshop support from the Michigan Center for Theoretical Physics. AE thanks
NSF grant AST-0708150.

Funding for the creation and distribution of the SDSS Archive has been provided by the Alfred P. Sloan 
Foundation, the Participating Institutions, the National Aeronautics and Space Administration, the National 
Science Foundation, the U.S. Department of Energy, the Japanese Monbukagakusho, and the Max Planck Society. 
The SDSS Web site is http://www.sdss.org/.

The SDSS is managed by the Astrophysical Research Consortium (ARC) for the Participating Institutions. 
The Participating Institutions are The University of Chicago, Fermilab, the Institute for Advanced Study, the 
Japan Participation Group, The Johns Hopkins University, the Korean Scientist Group, Los Alamos National Laboratory, 
the Max-Planck-Institute for Astronomy (MPIA), the Max-Planck-Institute for Astrophysics (MPA), New Mexico 
State University, University of Pittsburgh, University of Portsmouth, Princeton University, the United States 
Naval Observatory, and the University of Washington.



\newcommand\AAA[3]{{A\& A} {\bf #1}, #2 (#3)}
\newcommand\PhysRep[3]{{Physics Reports} {\bf #1}, #2 (#3)}
\newcommand\ApJ[3]{ {ApJ} {\bf #1}, #2 (#3) }
\newcommand\PhysRevD[3]{ {Phys. Rev. D} {\bf #1}, #2 (#3) }
\newcommand\PhysRevLet[3]{ {Physics Review Letters} {\bf #1}, #2 (#3) }
\newcommand\MNRAS[3]{{MNRAS} {\bf #1}, #2 (#3)}
\newcommand\PhysLet[3]{{Physics Letters} {\bf B#1}, #2 (#3)}
\newcommand\AJ[3]{ {AJ} {\bf #1}, #2 (#3) }
\newcommand\aph{astro-ph/}
\newcommand\AREVAA[3]{{Ann. Rev. A.\& A.} {\bf #1}, #2 (#3)}

\appendix

\section{A Maximum Likelihood Derivation of Matched Filter Richness Estimators}
\label{app:maxlkhd}

Here, we derive equation \ref{eq:mfrichness} using a maximum likelihood approach, focusing
first in the case where the filters $u(\bx|\lambda)$ are richness independent.
The derivation is as follows: we 
pixelize the observable space $\bx$ into infinitesimal pixels of ``volume'' $\Delta \bx$ such that every pixel contains
at most one galaxy.  The likelihood that a given galaxy realization occurs is simply
\begin{equation}
\lk \propto \prod_{occupied} (\lambda u+b)\Delta\bx \prod_{empty} (1-(\lambda u+b)\Delta\bx)
\label{eq:lkhd}
\end{equation}
where the first product is over all occupied pixels, while the second product is over all empty pixels.  We have neglected
terms that do not depend on $\lambda$ as they will not contribute to the maximum likelihood richness estimator.
Setting $\partial \ln \lk/\partial \lambda = 0$, and taking the limit $\Delta \bx \rightarrow 0$ we find that the maximum 
likelihood richness estimator $\hat \lambda_{ML}$ is given by the solution to
\begin{equation}
1 = \sum \frac{u}{\lambda u+b}
\end{equation}
where the sum is over all galaxies in the cluster field.  This expression is identical to our naive richness estimator 
from equation \ref{eq:mfrichness}.  

We wish to briefly consider how richness dependent filters $u(\bx|\lambda)$ affect the maximum 
likelihood richness estimator.  To do this, we go back to equation \ref{eq:lkhd}. Taking the derivative of the 
log-likelihood with respect to $\lambda$ and setting it to zero we find that the generalization of equation 
\ref{eq:mfrichness} is given by
\begin{equation}
1+ \int d\bx\ \lambda \frac{\partial u}{\partial \lambda} = \sum \frac{ u+\lambda(\partial u/\partial \lambda) }{\lambda u +b}.
\end{equation}
We emphasize that the integral over $\bx$ and the derivative $\partial/\partial\lambda$ do {\it not} always commute.  Indeed,
consider the approach taken in this paper, in which $u$ is taken to have a finite spatial extent of radius $R_c$, which is itself
linked to richness via equation \ref{eq:radius}.  The fact that $u$ is zero for $R>R_c(\lambda)$ implies that the integration region
for $u$ is $\lambda$ dependent, and thus the integral and derivative signs do not commute.

To assess the impact of a richness dependent profile, we consider here a simple isothermal filter
$u(R|\lambda)=1/R_c$, where $R_c(\lambda)$ is given by equation \ref{eq:radius}.\footnote{The two dimensional
density profile is, of course, $\Sigma(R)\propto 1/R$, but the radial probability density is $u(R)=2\pi R\Sigma(R) = 1/R_c$.}
For this filter, we have then
\begin{equation}
\lambda \frac{\partial u}{\partial \lambda} = \lambda\frac{\partial u}{\partial R_c} \frac{\partial R_c}{\partial \lambda} = - \alpha u
\end{equation}
where $\alpha$ is the slope of the radius-richness relation in equation \ref{eq:radius}.   Our expression for the maximum
likelihood richness estimator becomes
\begin{equation}
(1-\alpha) = \sum_{R<R_c(\lambda)}  \frac{ (1-\alpha)u }{\lambda u+b}.
\end{equation}
We see that the $1-\alpha$ prefactors cancel on both side of the equation, and thus our final expression for the maximum
likelihood richness estimator for $\lambda$ is still given by equation \ref{eq:richness1}, even though $u$ is explicitly
richness dependent.  This suggests that our naive estimator is in general very close to the true maximum likelihood 
estimator.  We defer a detailed study of whether the more complicated structure of the true maximum likelihood richness estimator
for more elaborate cluster profiles can lead to a significant improvement over the naive richness estimator from equation \ref{eq:richness1}
to future work.


\end{document}